\documentclass[prb,showpacs,preprintnumbers,twocolumn,amsmath,amssymb,superscriptaddress]{revtex4-1}

\usepackage[T1]{fontenc}
\usepackage[applemac]{inputenc}
\usepackage[english]{babel}
\usepackage{graphicx}
\usepackage{amssymb}
\usepackage{amsmath}
\usepackage{overpic}
\usepackage{color}

\usepackage{overpic}
\usepackage{mathtools}
\usepackage{braket}

\newcommand{\be}{\begin{equation}}
\newcommand{\ee}{\end{equation}}
\newcommand{\bea}{\begin{eqnarray}}
\newcommand{\eea}{\end{eqnarray}}

\def\br{{\bf r}}

\def\nn{\nonumber}
\def\lb{\label}
\def\pref#1{(\ref{#1})}

\newcount\bozza \bozza=0
\ifnum\bozza=1
\newdimen\shift \shift=-2truecm
\def\lb#1{%
{\label{#1}\rlap{\kern\shift{$\scriptstyle#1$}}}}
\else\def\lb#1{\label{#1}} \fi

\usepackage{feynmp}

\usepackage{ifpdf}
\ifpdf
\fi

\makeatletter
\def\endfmffile{%
  \fmfcmd{\p@rcent\space the end.^^J%
          end.^^J%
          endinput;}%
  \if@fmfio
    \immediate\closeout\@outfmf
  \fi
  \IfFileExists{\thefmffile.mp}{\immediate\write18{mpost \thefmffile}}{}
  \let\thefmffile\relax
}
\makeatother

\begin{document}
\title{Disordered XY model: effective medium theory and beyond}

\author{I.Maccari}
\affiliation{ISC-CNR and Department of Physics, Sapienza University of Rome, P.le A. Moro 2, 00185 Rome, Italy}

\author{L.Benfatto}
\affiliation{ISC-CNR and Department of Physics, Sapienza University of Rome, P.le A. Moro 2, 00185 Rome, Italy}

\author{C. Castellani}
\affiliation{Department of Physics, Sapienza University of Rome, P.le A. Moro 2, 00185 Rome, Italy}

\date{\today}%

\begin{abstract}
We study the effect of uncorrelated random disorder on the temperature dependence of the superfluid stiffness in the two-dimensional classical $XY$ model. By means of a perturbative expansion in the disorder potential, equivalent to the $T$-matrix approximation, we provide an extension of the effective-medium-theory result able to describe the low-temperature stiffness, and its separate diamagnetic and paramagnetic contributions. These analytical results provide an excellent description of the Monte Carlo simulations for two prototype examples of uncorrelated disorder. Our findings offer an interesting perspective on the effects of quenched disorder on longitudinal phase fluctuations in two-dimensional superfluid systems.
\end{abstract}

\maketitle

\section*{Introduction}

Despite its original formulation in terms of planar spins, the $XY$ model has been extensively investigated in the literature in the context of superconducting (SC) systems, which belong to the same universality class. From one side, the quantum $XY$ model properly describes the Josephson-like interactions between SC grains in artificial arrays\cite{fazio_review01}. From the other side, even for homogeneous superconductors  phase fluctuations at low temperature can be effectively described by coarse-grained $XY$-like models\cite{depalo_prb99,paramekanti_prb00,sharapov_prb01,benfatto_prb01,benfatto_prb04}. Within this context the equivalent of the Heisenberg interaction $J$ for spins becomes the energy scale connected to phase fluctuations, i.e. the superfluid superfluid stiffness $J_s=n_s^{2D}/4m$, where $n_s^{2D}=n_s^{3d}d$ is the effective two-dimensional superfluid density, and the length scale $d$ is the smallest between  the SC coherence length and the sample thickness.  In the specific case of quasi-2D systems, as thin SC films, the 2D $XY$ model allows also for a proper description of the topological phase transition due to the unbinding of vortex-like excitations, as described by the Berezinskii-Kosterlitz-Thouless theory\cite{bkt,bkt1,bkt2}. 

While in conventional clean superconductors $J$ is a much larger energy scale than the SC gap $\Delta$, making phase-fluctuation effects usually irrelevant, the suppression of $J$ in disordered superconductors or in unconventional ones brings back the issue of their possible role. In particular,  it has been proven  experimentally\cite{sacepe_10,mondal_prl11,chand_prb12,sherman2012,noat_prb13,brun_review17} that thin films of conventional superconductors 
at the verge of the superconductor-insulator transition (SIT) display a finite gap $\Delta$ above $T_c$, suggesting  that the phase transition is driven by  phase coherence more than pairing\cite{trivedi_prb01,dubi_nat07,ioffe,nandini_natphys11,seibold_prl12,lemarie_prb13}. 
Moreover, at strong disorder the SC ground state itself shows an emergent inhomogeneity\cite{sacepe_natphys11,mondal_prl11,pratap_13,noat_prb13,roditchev_natphys14,leridon_prb16},
 leading to a granular SC landscape. These findings suggest a mapping of the SC problem into an effective disordered bosonic system\cite{Anderson,MaLee}, whose phase degrees of freedom are conveniently described by quantum disordered $XY$ models. The underlying inhomogeneity triggers interesting effects, as e.g. the contribution of longitudinal quantum phase modes to the anomalous sub-gap optical absorption\cite{stroud_prb00,stroud_prb03,cea2014,swanson2014,pracht2017} as observed near the SIT
 \cite{armitage_prb07,driessen12,driessen13,frydman_natphys15,samuely15,armitage15,scheffler16}
or in films of nanoparticles\cite{bachar_jltp14,practh_prb16,pracht2017}. The non-trivial space structure of disorder can also have remarkable effects on the behavior of transverse (vortex-like) fluctuations in the classical limit, as we have recently investigated by means of Monte Carlo simulations\cite{BroadeningBKT,UnivBKT}. Indeed, by modelling the emergent granularity of the SC landscape with space-correlated inhomogeneous local couplings, we have shown that the anomalous nucleation of  vortices in the bad SC regions can lead to a  substantial smearing of the BKT superfluid-stiffness jump at the transition, in agreement with the systematically broadened jumps observed
experimentally both in thin films of conventional\cite{armitage_prb11,kamlapure_apl10,mondal_bkt_prl11,yazdani_prl13,yong_prb13} and unconventional\cite{lemberger_natphys07,lemberger_prb12,popovic_prb16}
superconductors. Apart from the smearing of the BKT jump, in Ref.\ \cite{BroadeningBKT} it has been observed that disorder can affect the low-temperature behavior of the superfluid stiffness in a non-trivial way, depending both on the variance of the disorder probability distribution and on its spatial correlations. 
These findings call for a deeper investigation of the general role of low-temperature phase fluctuations in classical disordered $XY$ models. Here we address this problem by combining Monte Carlo simulations with an analytical diagrammatic expansion. We start with the $XY$ Hamiltonian:
\be
\lb{H}
H_{XY}= - \sum_{i, \mu=\hat{x}, \hat{y}} J^{\mu}_i \cos(\theta_i - \theta_{i+\mu}),
\ee
where disorder is encoded in the local couplings $J^{\mu}_i$. We consider two possible types of spatially uncorrelated disorder. The first one is the case of a Gaussian distribution, that is usually employed to mimic relatively weak fluctuations of the local stiffnesses around a given mean value. The second one is the diluted model, where a fraction $p$ of the couplings is taken equal to zero, mimicking the local suppression of the Josephson coupling between neighbouring SC regions due to disorder. Despite being a model without specific spatial correlations for the disorder, its SC properties are nonetheless ultimately dominated by the global phenomenon of percolation\cite{kirkpatrick}. By means of Monte Carlo simulations we compute the temperature dependence of the superfluid stiffness for increasing disorder. These results are compared with the analytical derivation of the low-temperature stiffness obtained from the calculation of the self-energy corrections to the phase propagator due to disorder. By resumming in the disorder, within the on-site $T$-matrix scheme\cite{Mahan}, we derive a self-consistence equation for the stiffness that is formally equivalent to the results obtained in the effective-medium-theory  approximation\cite{kirkpatrick,stroud_prb00} (EMA). This approach allows us to generalize the EMA result to include finite-temperature corrections. The derived analytical formulas are in excellent agreement with the Monte Carlo simulations, and allow one to capture the role played by different disorder models for the thermal activation of longitudinal phase fluctuations.\\
The plan of the paper is the following. In Sec. \text{I}, we briefly review the standard results expected for the clean $XY$ model. In Sec. II, we consider the disordered case. After reviewing the equivalence with the random-resistor-network problem, we first solve the zero-temperature case (Sec. IIA), recovering the analogy between the $T$-matrix resummation in the disorder potential and the EMA  not only for the global stiffness but also for the separate diamagnetic and paramagnetic contributions. In Sec. IIB, we use the same strategy to derive a modified EMA equation able to include the finite-temperature corrections to the stiffness in the presence of disorder. In Sec. \text{III}, we compare these analytical expressions with the numerical results obtained by means of Monte Carlo simulations.  Sec. IV contains the closing discussion and remarks.

\section{Temperature dependence of the stiffness in the clean case}
Before discussing the role of disorder, we start by briefly recalling the standard results expected for the clean $XY$ model. We will thus consider the model \pref{H} on a $L\times L=N$ two-dimensional square lattice with homogeneous couplings $J^{\mu}_i=J$.  The superfluid stiffness is the response to a transverse gauge field, that can be minimally coupled to the SC phase by the replacement:
\be
\lb{H_hom}
H_{XY}= - J\sum_{i, \mu=\hat{x}, \hat{y}}  \cos(\theta_i - \theta_{i+\mu}+A_{\mu}).
\ee
As usual, due to the periodic boundary conditions, applying a constant field ${\bf A}$ along say the $x$ direction is equivalent to consider twisted boundary conditions for the phase, with a total flux $\phi=A_xL$ (in units of $\Phi_0=hc/2e$ for the SC case) through the sample. The current density is defined as $I_x=-N^{-1}\partial H/\partial A_x$, so that at leading order in $A_x$ one has
\bea
I_x&=&-\frac{J}{N}\sum_{i}\sin(\theta_i - \theta_{i+\hat{x}}+A_x)\simeq\nn\\
&\simeq&-\frac{J}{N}
\sum_{i} \Big[ \sin(\theta_i - \theta_{i+\hat{x}})+\cos(\theta_i - \theta_{i+\hat{x}})A_x \Big],
\eea
where the first term defines the paramagnetic current and the second one is the diamagnetic response.
By computing the average current by linear response in $A_x$ and defining the stiffness as $J_s=-\langle I_x\rangle /A_x$ one has:
\bea
\lb{Jsc}
J_s&=&J_d-J_p\\
\lb{jdc}
J_d&=& \frac{J}{N}{\braket{ \sum_i \cos(\theta_i - \theta_{i+\hat x}) }}; \\
\lb{jpc}
J_p&=& \frac{J^2}{NT}{\braket{(\sum_i   \sin(\theta_i - \theta_{i+\hat x}) )^2 }},
\eea
where $T$ is the temperature. 
As usual, the paramagnetic contribution $J_p$ is the correlation function for the paramagnetic current, while the diamagnetic contribution $J_d$ coincides in this case with the average energy density along the $x$ direction. To estimate their contribution to the thermal suppression of the superfluid stiffness at low temperature we can approximate  the phase difference between neighboring  sites $\theta_{i+\mu}-\theta_i$ with a continuum gradient $\theta_{i+\mu}-\theta_{i}\approx \partial_{\mu} \theta(\br_i)$, where we set the lattice spacing $a=1$. By expanding also the cosine of Eq.\ \pref{H_hom} at leading order in the phase gradient we end up with a Gaussian model accounting only for spin-wave like longitudinal phase fluctuations:
\be
\lb{gauss}
H_{XY}\simeq \frac{J}{2} \int d{\bf r} \big(\nabla\theta({\bf r})\big)^2.
\ee
The approximation \pref{gauss} allows one to perform analytically the averages in Eq.s\ \pref{jdc}-\pref{jpc}. For the diamagnetic term one immediately finds:
\bea
\lb{jdc_app}
J_d&\simeq& J-\frac{J}{2N}\braket{\int  d{\bf r} \big(\partial_x \theta({\bf r})\big)^2}\simeq J-\frac{T}{2d}.
\eea
where we used that fact that for the Gaussian model \pref{gauss} $\braket{\int  d{\bf r} \big(\partial_x \theta({\bf r})\big)^2}=T/dJ$, with $d=2$ the space dimension.
For the paramagnetic contribution, we expand the sine function in powers of the phase gradient. The correlation function \pref{jpc} amounts then to the sum of processes where an odd number of phase modes (phasons) are excited by the electromagnetic field. However, in the clean case the first term, proportional to $\propto\braket{\int d\br d\br' \partial_x \theta(\br) \partial_x \theta(\br')}$,  vanishes because of periodic boundary conditions. The next non-zero contribution is then a three-phasons processes, which reads explicitly:
\bea
J_p&\simeq& \frac{J^2}{TN} \braket{\int  d\br d\br'  \frac{1}{6} \big(\partial_x \theta({\bf r})\big)^3  \frac{1}{6}\big(\partial_x \theta({\bf r}')\big)^3}=\nn\\
\lb{jpc_app}
&=&J^2\frac{T^2}{6 d^3}.
\eea

\begin{figure}[t]
\includegraphics[width=\linewidth]{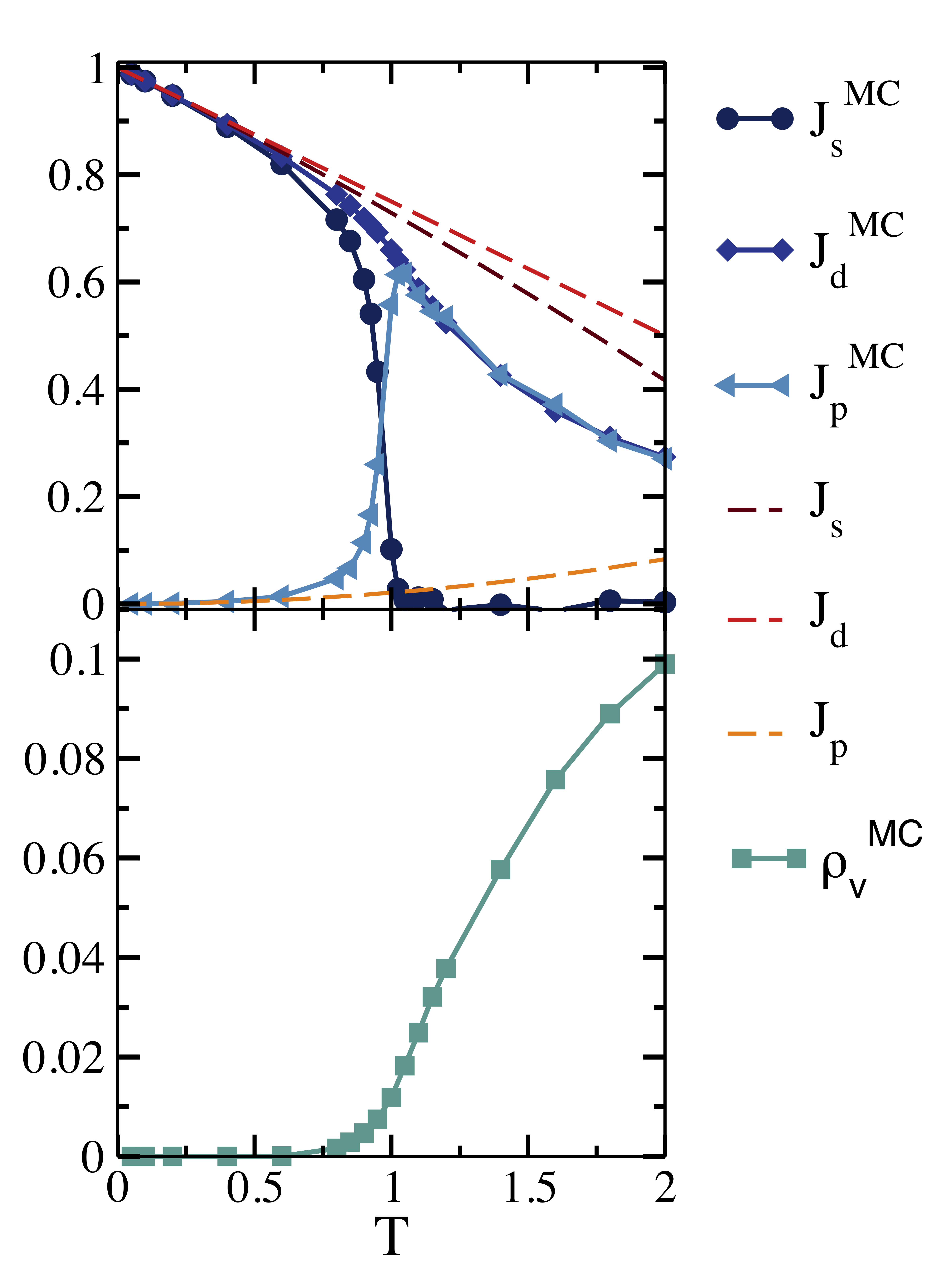}
\caption{(Color online) Upper panel: comparison between the Monte Carlo results and the low-temperature analytical expressions for the three response functions $J_s$, $J_d$ and $J_p$ based on Eq.s \pref{jdc_app} and \pref{jpc_app}. Lower panel: vortex density of the system as function of the temperature. The linear size of the system simulated is $L=256$. 
}
\label{fig1}
\end{figure}
The analytical expressions \pref{jdc_app}-\pref{jpc_app} perfectly reproduce the low-temperature Monte Carlo simulations shown in Fig.\ \ref{fig1}. In particular the diamagnetic suppression \pref{jdc_app} is the main source of temperature dependence of the stiffness up to $T\simeq 0.6$. As the temperature increases two effects come into play. From one side higher-order terms in the phase gradient should be included, and more importantly vortex-antivortex pairs start to form and unbind at the BKT transition, as shown in the lower panel of Fig.\ \ref{fig1}. The latter effect appears predominantly in the paramagnetic contribution, which includes large-distance current correlations, while the diamagnetic one is predominantly local. As a consequence $J_p$ sharply increases approaching $T_c$, causing the rapid downturn of the superfluid stiffness that is the signature of the universal BKT jump in  simulations on a finite-size system. 
%

It is worth noting that the general expressions \pref{jdc_app} and \pref{jpc_app} of the diamagnetic and paramagnetic temperature corrections in powers of the phase modes hold also for the quantum case. However, in this case the average value of the phase gradient should be computed using a quantum phase model, including the frequency dependence of the phase fluctuations\cite{depalo_prb99,paramekanti_prb00,sharapov_prb01,benfatto_prb01,benfatto_prb04}. The main consequence for the present discussion is that below a crossover temperature $T_{cl}$ the classical (thermal) corrections to the superfluid stiffness discussed so far turn into quantum ones, leading finally to a $T=0$ value of the stiffness smaller than $J$ even for clean models. The exact form of the quantum corrections depends on the dynamics of the phase mode, that in turn is controlled by the presence of Coulomb interactions and dimensionality (see e.g. \cite{fazio_review01,benfatto_prb04}). In ordinary  SC systems the quantum phase modes are pushed to the plasma energy scale, so that $T_{cl}$ can be larger than the critical temperature  itself where quasiparticle excitations destroy the SC state, hindering the observation of phase-fluctuation effects on the superfluid stiffness. On the other hand, whenever the plasma energy scale is suppressed by correlations or disorder,  $T_{cl}$ can be considerably reduced making eventually classical phase-fluctuations corrections to the stiffness experimentally accessible. This possibility has been  discussed within the context of both unconventional superconductors\cite{benfatto_prb01} and 
strongly-disordered conventional ones\cite{mondal_prl11}. For this reason, the classical $XY$ model discussed in the present manuscript is not only interesting by itself, but it can be relevant also for applications to real systems.

\section{Disordered case}

Let us consider now the case where the local couplings $J^\mu_{i}$ of the model \pref{H} are random variables extracted with a probability distribution $P(J)$. In full analogy with the clean case, the superfluid stiffness is defined via diamagnetic and paramagnetic terms, which are now also averaged over disorder configurations:
\bea
\lb{jd}
J_d&=& \frac{1}{N}\overline{\braket{ \sum_i J^{\mu}_i \cos(\theta_i - \theta_{i+\mu}) }}; \\
\lb{jp}
J_p&=& \frac{1}{NT}\overline{\braket{(\sum_i J^{\mu}_i  \sin(\theta_i - \theta_{i+\mu}) )^2 }};
\eea
In the disordered case the $T=0$ value of the stiffness can be obtained by mapping the SC problem into the random-resistor network (RRN) one.
The starting point is again a Gaussian approximation for the cosine term in Eq.\ \pref{H}, so that
\begin{equation}
\lb{Hxy_LowT}
H_{XY} \simeq - \sum_{i, \mu=\hat{x}, \hat{y}}  J^{\mu}_i (1 - \frac{1}{2} (\theta_i - \theta_{i+\mu})^2).
\end{equation}
\noindent
The configuration of the phase variables can be obtained by the minimization of Eq.\ \pref{Hxy_LowT}, giving a set of equations:
\begin{equation}
\lb{jeqs}
\sum_{\mu=\pm \hat{x}, \hat{y} } J^{\mu}_i  (\theta_i - \theta_{i+\mu} ) =0.
\end{equation}
One can then recognize the analogy between Eq.s\ \pref{jeqs} and the Kirchhoff equations relating the value of the current $i^\mu_i$ between two nodes $i$ and $ i+\mu$ in terms of the local voltages $V_i$ and the link conductances $\sigma^\mu_i$
\begin{equation}
\lb{node}
\sum_{\mu=\pm \hat{x}, \hat{y} } i^{\mu}_i=\sum_{\mu=\pm \hat{x}, \hat{y} } \sigma^{\mu}_i (V_i - V_{i+ \mu})=0.
\end{equation}
The comparison between Eq.s \pref{jeqs} and \pref{node} establishes the equivalence between the local conductances $\sigma^\mu_i$ of the RRN and the local stiffnesses $J^{\mu}_i $ of the $XY$ model. This also means that finding the global phase stiffness $J_s$ is equivalent to determine the global conductance $\sigma$ of the RRN problem. A possible solution for $\sigma$ has been proposed long ago within the  Effective-Medium approximation (EMA) scheme (see \cite{kirkpatrick} and references therein).  The basic idea is that the inhomogeneous system can be mapped into a homogeneous one characterized by an effective value $\tilde \sigma$ of the conductance such that, on average, the presence of a single disordered link with $\sigma_i \neq \tilde \sigma $ has vanishing effects on the current and voltage distributions of the system. The effective conductance $\tilde\sigma$ is then obtained as solution of a self-consistent equation, that for a cubic lattice in $d$ dimensions reads:
\begin{equation}
\lb{RRN}
\sum_{i} P_i \frac{\sigma_i - \tilde \sigma}{ \sigma_i + (d-1)\tilde \sigma}=0,
\end{equation}
where $P_i$ is the probability for the occurrence of each possible $\sigma_i$ value. 
In this manuscript we will provide a derivation of Eq.\ \pref{RRN} for the superfluid stiffness based on resummed perturbation theory. 
Within our approach we will estimate not only the $T=0$ superfluid stiffness, but also its leading temperature dependence, giving an excellent description of Monte Carlo simulations. Moreover, we will derive the effect of disorder on the two separate diamagnetic and paramagnetic contributions, that can be relevant for the experiments. Indeed, thanks to the optical sum rule\cite{stroud_prb00} one knows that the extra paramagnetic suppression of the stiffness induced by disorder transfers into a 
finite-frequency optical absorption. The quantum version of this mechanism has been recently invoked to explain the extra sub-gap microwave absorption measured at low temperatures in strongly disordered superconductors (see e.g. Ref.s \onlinecite{swanson2014,cea2014,seibold_prb17} and references therein). As mentioned below, the analogous classical effect investigated here can be relevant in real materials above the crossover temperature $T_{cl}$. 

\subsection{Effective medium theory for the XY model at $T=0$}

As explained in Sec. I, at low temperature, where the topological phase excitations (vortices) still play no role,  a continuum approximation for the model \pref{H} allows one to easily describe the longitudinal phase fluctuations. In the disordered case we will follow the same strategy, by implementing also the basic idea underlying the EMA. We then introduce an homogeneous effective stiffness $\tilde J$ and we will require that adding a single impurity with a local stiffness different from $\tilde J$  will have no overall effect on the system. Let us, thus, write the XY Hamiltonian as the sum of the two terms:
\be
\lb{H0Hi}
H = H_0 + H_i= \frac{\tilde{J}}{2} \int d{\bf r} \big(\nabla\theta({\bf r})\big)^2 + \frac{\delta J_i}{2} \big(\nabla \theta({\bf r_i})\big)^2 ,
\ee
where we put $J_i= \tilde{J} + \delta J_i$. The second term of Eq.\ \pref{H0Hi} can be seen as a perturbation with respect to $H_0$. We can then compute the self-energy correction $\Sigma_i$ to the bare Green's function $G_0=\braket{\theta_{{\bf k }}\theta_{-{\bf k}}}_{H_0}$ due to the presence of the impurity, so that the total Green's function $G({\bf k})= \braket{\theta_{{\bf k }}\theta_{-{\bf k}}}_{H_0+H_i}$ reads:
\be
\lb{Dyson}
G({\bf k})= G_0({\bf k}) + G_0({\bf k})\Sigma_{i}({\bf k})G_0({\bf k}).
\ee
Be expanding $e^{-\beta H_i}$ in power series we can compute $G(\mathbf{k})$ as,
\be
 G({\bf k})= \braket{\theta_{{\bf k }}\Big[ \sum_{n=0}^{\infty} \frac{(\beta)^n}{n!}\Big(-\frac{\delta J_i}{2}\Big)^n  (\nabla \theta ({\bf r}_i))^{2n}\Big]\theta_{-{\bf k}}}_{H_0}
\lb{G}
\ee
\noindent
The first therm ($n=0$) is nothing but the bare Green's function $G_0({\bf k})= {T}/{\tilde{J} \mathbf{k}^2}$. The remaining terms can be computed by means of Wick's theorem. For example the second term ($n=1$) reads:
\bea
& &G_{1}({\bf k})= \braket{\theta_{{\bf k }}\Big[ \frac{-\beta \delta J_i}{2}(\nabla \theta ({\bf r}_i))^{2}\Big]\theta_{-{\bf k}}}_{H_0}\nn\\
&=& \braket{\theta_{{\bf k }}\Big[ \frac{-\beta \delta J_i}{2N}\sum_{ \mathbf{q_1},\mathbf{q_2}} e^{i(\mathbf{q_1}+ \mathbf{q_2}) \cdot \mathbf{r_i}} (i \mathbf{q_1})(i \mathbf{q_2}) \theta_{\mathbf{q_1}} \theta_{\mathbf{q_2}}\Big]\theta_{-{\bf k}}}_{H_0}=\nn\\
&=&G_0({\bf k}) \Big( - \frac{\delta J_i}{N} \, \frac{\mathbf{k}^2 }{T} \Big) G_0({\bf k})= G_0({\bf k}) \Sigma_i^{(1)}({\bf k}) G_0({\bf k}),
\label{sigma1}
\label{G1}
\eea
where the prefactor $1/2$ canceled out with the diagram multiplicity. The factor $1/N$, due to the fact that  we are considering  one single impurity, will be omitted in what follows. Indeed, as soon as one considers the original model with $N$ non-interacting possible impurities, the sum over all impurities cancels out the $N$ prefactor. 
%
\begin{figure*}[t!]
\includegraphics[width=\linewidth]{{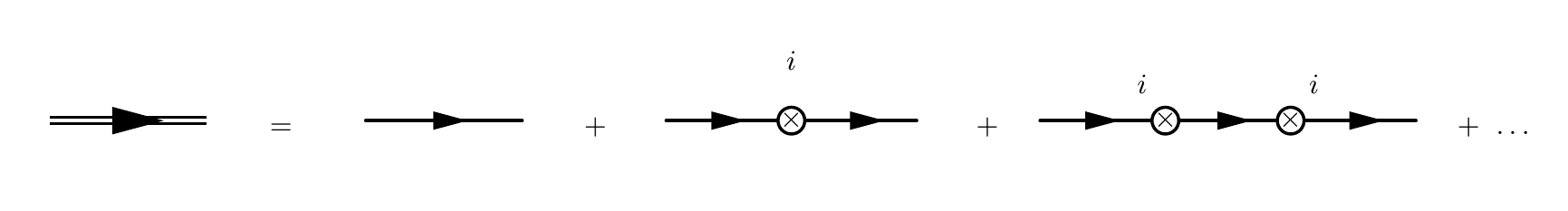}}
\caption{Diagrammatic representation of the local self-energy corrections to the bare Green's function at $T=0$. Here the solid line represents the bare Green's function $G_0$, the double line the dressed one $G$ and each cross account for a single-impurity scattering, contributing with a factor proportional to $\delta J_i$.}
\label{sum_green}
\end{figure*}
%
For higher order terms this procedure implements the usual $T$-Matrix approximation\cite{Mahan}, where only non-crossing diagrams with multiple scattering events by different impurities are included. It is then easy to verify that the $n$-th term of the expansion reads:

\be
\begin{split}
G_{n}({\bf k})&= G_0({\bf k}) \Big[ (-1)^n \frac{\mathbf{k}^2 }{T} \, \Big(\frac{\delta J_i}{d \tilde{J}} \Big)^n \, d \tilde{J} \Big] G_0({\bf k})=\\&= G_0({\bf k}) \Sigma_i^{(n)}({\bf k}) G_0({\bf k}).
\end{split}
\label{sigman}
\ee
\noindent
Therefore, we can write the full local single-impurity self-energy as:
\be
\Sigma_i({\bf k})= \sum_{n=1}^{\infty} \Sigma_i^{(n)}({\bf k})= \sum_{n=1}^{\infty}  (-1)^n \frac{\mathbf{k}^2 }{T} \, \Big(\frac{\delta J_i}{d \tilde{J}} \Big)^n \, d \tilde{J},
\ee
whose diagrammatic representation is shown in Fig.\ref{sum_green}.
The irrelevance of the single-impurity perturbation on the physical responses of the system translates, in this approach, in the request of a vanishing local self-energy at all orders in the perturbation. This in turn is satisfied if:
\be
\lb{request}
\sum_i P_i \sum_{n=1}^{\infty}  (-1)^n \Big(\frac{\delta J_i}{d \tilde{J}} \Big)^n=0,
\ee
where we included also the average over all the possible values of $\delta J_i$, extracted from the probability distribution $P_i\equiv P(J_i)$. Since $\delta J_i =\tilde{J} - J_i$, we can rewrite Eq.\eqref{request} as:
\be
\sum_i P_i \Big[ \sum_{n=0}^{\infty}  (-1)^n \Big(\frac{\tilde{J}-  J_i}{d \tilde{J}} \Big)^n -1 \Big]=0,
\ee
which, after simple algebra, is equivalent to:
\be
\sum_i P_i \Big[ \frac{ J_i - \tilde{J}}{J_i + (d-1)\tilde{J}}\Big]=0.
\label{EMT}
\ee

By direct comparison with \eqref{RRN} we immediately see that $\tilde J$ satisfies the same equation of the effective conductance of the RRN model within EMA. While this was expected on the basis of the formal analogy between the two problems encoded in Eq.s \pref{jeqs} and \pref{node}, our derivation of the equivalence  between the EMA equation and the perturbative expansion in the disorder potential is a new result, which brings interesting consequences. Following the same procedure we can compute separately the zero-temperature values of the diamagnetic $J_d(T=0)$ and paramagnetic $J_p(T=0)$ contributions,   in order to understand how the diagrammatic expansion in $H_i$ affects the clean-limit results.
We start by gradient expansion of Eq. \eqref{jd} and \eqref{jp}. For the diamagnetic contribution the first-order correction in the cosine expansion already gives a finite-$T$ correction, in full analogy with Eq.\ \pref{jdc_app} above, so we simply have :
\begin{equation} 
J_d (T=0) \simeq \frac{1}{N}\overline{\braket{\int d \mathbf{r} J(\mathbf{r})}} _{H_0+ H_i} =\bar{J},
\lb{Jd_T0}
\end{equation}
showing that at all orders in the perturbing potential the zero-temperature diamagnetic response coincides with  the mean value $\bar{J}$ of the couplings. For the paramagnetic term disorder makes different from zero the single-phason process discussed in the previous section, that then reads:
\begin{equation} 
J_p \simeq   \frac{1}{NT} \overline{ \braket{ \Big[\int d\mathbf{r} \, J(\mathbf{r}) \Big(\nabla_x \theta(\mathbf{r}) \Big) \int d\mathbf{r'} \, J(\mathbf{r'}) \Big(\nabla_x \theta(\mathbf{r'}) \Big) }}_{H_0+ H_i} .
\label{Jp_T0}
\end{equation}
By using the fact that $J({\bf r})=\tilde J$ everywhere except in the single impurity site where $J_i=\tilde J+\delta J_i$, and that the contribution proportional to the homogeneus stiffness $\tilde J$ vanishes because of periodic boundary conditions, we immediately get
\begin{equation}
\begin{split}
J_p &\simeq  \frac{1}{NT} \overline{ \braket{ \Big[(\delta J_i)^2 \Big(\nabla_x \theta(\mathbf{r_i}) \Big)^2\Big] }}_{H_0+ H_i}. 
\end{split}
\label{Jp_T0_3}
\end{equation}
Now, by omitting as before one overall prefactor $1/N$, we proceed with the perturbation expansion in $H_i$. The first term ($n=0$) is simply:
\be
\begin{split}
{J_{p}}_{(n=0)}&=\frac{(\delta J_i)^2}{NT} \braket{\sum_{{\bf q}_1, {\bf q}_2 } e^{i({\bf q}_1+ {\bf q}_2) \cdot \mathbf{r}_i} (i \mathbf{q}^x_1) (i \mathbf{q}^x_2) \theta_{{\bf q}_1} \theta_{{\bf q}_2}}_{H_0}=\\ &= \frac{(\delta J_i)^2}{d \tilde{J}}.
\end{split}
\lb{jplead}
\ee
The second term ($n=1$) will be instead:
\be
\begin{split}
{J_{p}}_{(n=1)}= \frac{-(\delta J_i)^3}{2N^2T}&\langle \sum_{{\bf q}_1, {\bf q}_2 }\sum_{{\bf k}_1, {\bf k}_2 }  e^{i({\bf q}_1+ {\bf q}_2 +{\bf k}_1+ {\bf k}_2) \cdot \mathbf{r}_i}\cdot \\ & \cdot \mathbf{q}^x_1 \mathbf{q}^x_2  \mathbf{k}_1 \mathbf{k}_2  \theta_{{\bf q}_1} \theta_{{\bf q}_2}  \theta_{{\bf k}_1} \theta_{{\bf k}_2}  \rangle_{H_0},
\end{split}
\ee
which after the contractions reads:
\be
{J_{p}}_{(n=1)}=-\frac{(\delta J_i)^3}{d^2 \tilde{J}^2}.
\ee
By proceeding analogously at all orders in the perturbative expansion one can write the zero-temperature value of the paramagnetic term as:
\be
\begin{split}
&J_p(T=0)= \sum_i P_i  \Big [d \tilde{J} \sum_{n=2}^{\infty} (-1)^n \Big(\frac{\delta J_i}{d \tilde{J}}\Big)^ n\Big]=\\ &= \sum_i P_i  \Bigg\{ d \tilde{J} \Big[ \Big( \sum_{n=1}^{\infty} (-1)^n \Big(\frac{\delta J_i}{d \tilde{J}}\Big)^ n \Big) -\Big( -\frac{\delta J_i}{d \tilde{J}} \Big) \Big] \Bigg\}=\\ &= \sum_i P_i \Big [ d \tilde{J} \, \frac{\delta J_i}{d \tilde{J}} \Big] = \sum_i P_i (J_i - \tilde{J})=\bar J-\tilde J,
\end{split}
\lb{jp0}
\ee
where, in the last step, we used the result of Eq.\pref{request}. Eq.s\ \pref{Jd_T0} and \pref{jp0} clearly satisfy the general relation \eqref{Jsc}, as expected. On the other hand, their separate evaluation helps understanding the different role of disorder on the two terms. Indeed, while the diamagnetic term \pref{Jd_T0} is a measure of the average disorder distribution, the paramagnetic one is a measure of its variance, as one immediately see from the leading correction \pref{jplead}. This results will help us explaining the difference between Gaussian and diluted disorder in Sec. III.

\subsection{Effective medium theory for the XY model up to linear terms in $T$}

Let us now extend the $T=0$ results in order to estimate the leading temperature corrections to $J_s$, $J_d$ and $J_p$. To this aim we need to consider that each $G_0$ carries a power of $T$. By power counting it is then clear that at finite $T$ it is crucial to retain in the disorder Hamiltonian $H_i$ also the quartic term in the expansion of the cosine:
\be
H_i=H_i^{(2)}+H_i^{(4)}=\frac{ \delta J_i}{2}  (\nabla \theta ({\bf r}_i))^{2} -  \frac{J_i}{4!}\sum_{\mu=\hat{x},\hat{y}} (\nabla_{\mu} \theta ({\bf r_i}))^{4} 
\lb{h24}
\ee
From the diagrammatic point of view, the quartic $H_i^{(4)}$ term in $\nabla \theta(\mathbf{r_i})$ introduces a 4-legs vertex in the phase field, whose combination with the 2-legs one in $H_i^{(2)}$ complicates the calculation of the Green's function, that will be carried out along the same lines of Eq.\ \pref{G}. The easiest way to handle this problem is to follow the same logic of the zero-temperature case, as summarized in Fig.\ \ref{sum_greenT}. When computing the full Green's function (double line) we only sum up non-crossing diagrams with multiple scattering by a single impurity site. However, we will replace the $\delta J_i$ (crossed circle) with its finite-temperature value $\delta J_i(T)$:

\begin{figure*}[t]
\includegraphics[width=\linewidth]{{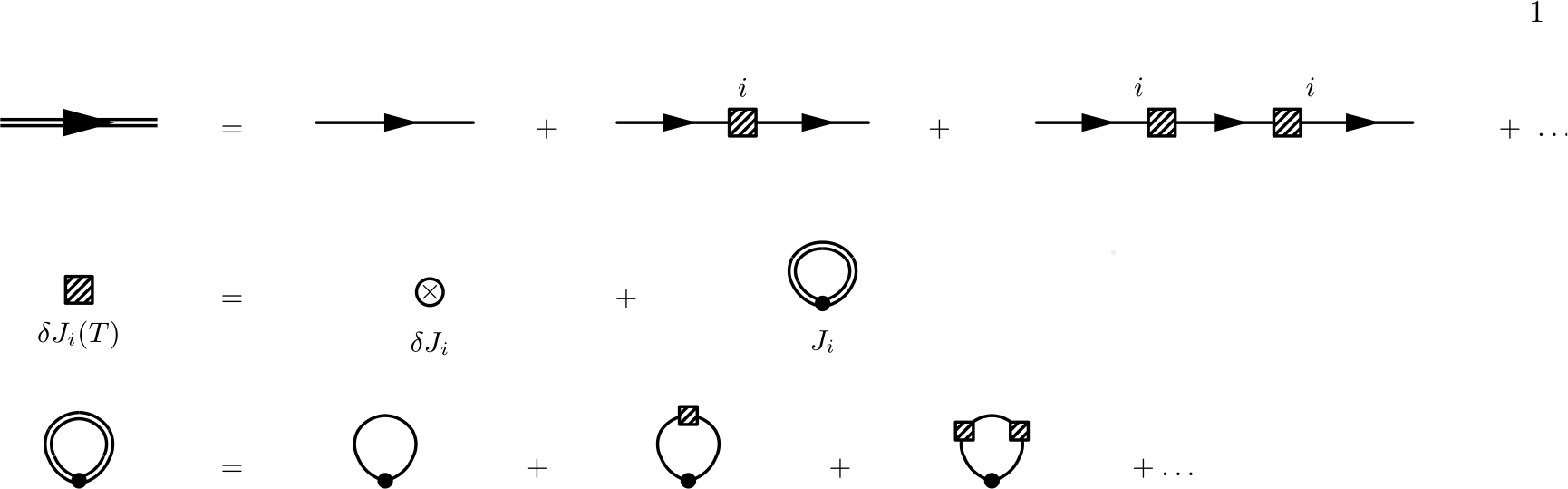}}
\caption{Diagrammatic representation of the local self-energy corrections to the bare Green's function up to the linear terms in temperature.  In the diagrammatic expansion, each arrow stays for a bare Green's function, while the shaded square accounts for two different impurity-scattering contributions. As shown in the second line, indeed, it corresponds to the sum of the zero-temperature correction, proportional to $\delta J_i$, and the linear-temperature one, proportional to $J_i$, arising from the 4-legs vertex in $H_i^{(4)}$. The closed loop appearing in the second line accounts self-consistently for all the local self-energy corrections, as shown in the last line. As a matter of fact, however, since we are considering the self-energy corrections up to the linear terms in temperature, in our calculation we will consider the closed loop corrected by only the zero-temperature contribution, i.e. we will replace the dashed square with the bare circle $\delta J_i$. }
\label{sum_greenT}
\end{figure*}
\be
\delta J_i \to \delta J_i(T).
\ee
In this way, $\delta J_i(T)$ accounts for scattering events described both by the $H_i^{(2)}$ 2-legs vertex insertions and the $H_i^{(4)}$ 4-legs vertex insertion, which generates a loop diagram with the full Green's function, as shown in Fig.\ 
\ref{sum_greenT}. 
The first order of the new local self-energy is then obtained by the second line of Fig.\ref{sum_greenT}. The first term is the  zero-temperature contribution, already given in Eq.\eqref{G1}:
\be
G_{(n=1)}(\mathbf{k}, O(1))= G_0({\bf k}) \Big( - \delta J_i \, \frac{\mathbf{k}^2 }{T} \Big) G_0({\bf k})
\label{Gk1_O(1)}
\ee
The second term arises from the 4-legs vertex and it corresponds to a loop diagram, as shown in the second line of Fig.\ \ref{sum_greenT}:
\be
\lb{G1_T}
\begin{split}
&G_{(n=1)}(\mathbf{k}, O(T))= \\
&=\braket{\theta_{{\bf k }}\Big[ \frac{\beta}{4!}J_i \Big( \nabla_x \theta(\mathbf{r}_i)  \Big)^4+  \Big( \nabla_y \theta(\mathbf{r}_i) \Big)^4 \Big] \theta_{-{\bf k }}}_{H_0},
\end{split}
\ee
being, $J_i= \tilde J +  \delta J_i$. 
We can then rewrite Eq. \pref{G1_T} in momentum space as:
\be
\begin{split}
G_{(n=1)}(\mathbf{k}, O(T))&= 2 \langle \theta_{{\bf k }}\Big[ \frac{\beta J_i}{4!} \sum_{{\bf q}_1, {\bf q}_2, {\bf q}_3, {\bf q}_4 } e^{i ({\bf q}_1 + {\bf q}_2 + {\bf q}_3 + {\bf q}_4)\cdot \mathbf{r}_i}\cdot \\ &\cdot {\bf q}^x_1 {\bf q}^x_2 {\bf q}^x_3 {\bf q}^x_4 \theta_{{\bf q}_1}  \theta_{{\bf q}_2} \theta_{{\bf q}_3} \theta_{{\bf q}_4}\Big] \theta_{-{\bf k }}\rangle_{H_0},
\end{split}
\label{Gk1_O(T)}
\ee
%
where the prefactor $2$ accounts for the gradient along $y$ in Eq.\ \pref{G1_T}.
We then proceed with the Wick's theorem to end up with:
\be 
\begin{split}
&G_{(n=1)}(\mathbf{k}, O(T))=\\&= G_0(\mathbf{k}) \Big[ \beta J_i \frac{\mathbf{k}^2}{2d} \sum_{\mathbf{q}} \mathbf{q}^2 \braket{\theta_{\mathbf{q}} \theta{-\mathbf{q}}}_{H_0} \Big]  G_0(\mathbf{k}).
\end{split}
\lb{eqGin}
\ee
The quantity in square brackets in the above equation defines the Green's function in the loop, that at first order coincides with $G_0$. However,  higher order terms in the expansion of $H_i$ lead to the dressing of this loop by all the possible single-impurity scattering processes, as shown by the last line of Fig.\ \ref{sum_greenT}.  
The easiest way to sum them up is to replace the average over $H_0$ in Eq.\ \pref{eqGin} with an average over $H_0+H_i$. In addition, since this is already a finite-temperature correction, we can restrict ourself to the terms generated by $H_i^{(2)}$, as done for the $T=0$ case. We then have that
\be 
\begin{split}
&G_{(n=1)}(\mathbf{k}, O(T))=\\&= G_0(\mathbf{k}) \Big[ \beta J_i \frac{\mathbf{k}^2}{2d} \sum_{\mathbf{q}} \mathbf{q}^2 \braket{\theta_{\mathbf{q}} \theta{-\mathbf{q}}}_{H_0+H_i^{(2)}} \Big]  G_0(\mathbf{k})=\\
&=G_0(\mathbf{k}) \Big[ \beta J_i \frac{\mathbf{k}^2}{2d} \sum_{\mathbf{q}} \mathbf{q}^2 G_0(\mathbf{q}) \Big( \sum_{m=0}^{\infty} (-1)^m (\frac{\delta J_i }{d \tilde{J}})^m\Big) \Big]  G_0(\mathbf{k}).
\end{split}
\ee

%
Solving then the geometric series in the last line, one ends up with:
\be
G_{(n=1)}(\mathbf{k}, O(T))= G_0(\mathbf{k}) \Big[ \frac{\mathbf{k}^2}{2}  \frac{J_i }{d \tilde{J} + \delta J_i} \Big]  G_0(\mathbf{k}).
\label{Gk1_O(T)final}
\ee
Finally, putting together Eq.\eqref{Gk1_O(1)} and Eq.\eqref{Gk1_O(T)final} we obtain the explicit expression for the single-scattering diagram on the upper line of Fig.\ \ref{sum_greenT}:
\be 
\begin{split}
G_{(n=1)}(\mathbf{k})&=  G_0(\mathbf{k}) \Big[ - \frac{\mathbf{k^2}}{T} \Big( \delta J_i - \frac{T}{2} \frac{J_i}{(d-1)\tilde{J} + J_i}\Big) \Big]  G_0(\mathbf{k})=\\&=G_0(\mathbf{k}) \Big[ - \frac{\mathbf{k^2}}{T} \delta J_i (T) \Big]  G_0(\mathbf{k}), 
\end{split}
\ee
\noindent
which is the equivalent of Eq.\eqref{G1} up to linear terms in temperature. Thus, in perfect analogy with Eq.\eqref{request}, the new EMT equation reads:
 \be
\sum_i P_i \sum_{n=1}^{\infty}  (-1)^n \Big(\frac{\delta J_i(T)}{d \tilde{J}(T)} \Big)^n= \sum_i P_i \frac{-\delta J_i(T)}{d\tilde{J}(T) +\delta J_i(T)} =0,
\lb{new_EMT}
\ee
where $\delta J_i(T)$is given by:
\be
\delta J_i(T)=  \delta J_i - \frac{T}{2} \frac{J_i}{(d-1)\tilde{J}(T) + J_i}.
\lb{deltaJT}
\ee
 
The same strategy can now be used to compute the finite-temperature corrections to the diamagnetic and paramagnetic terms. For the diamagnetic response the leading dependence on temperature is given by the first term in the cosine expansion of Eq.\eqref{jd}, so that:

\begin{equation}
 J_d(T) =\bar{J} -\frac{1}{2}\overline{\braket{ \int d\mathbf{r} (\tilde{J} \Big(\nabla_x \theta(\mathbf{r})\Big)^2 + \delta J_i \Big( \nabla \theta_x(\mathbf{r}_i)\Big)^2}}_{H_0+ H_i^{(2)}}
\label{expandingJd}
\end{equation}
where we used again only the two-leg vertex of the impurity Hamiltonian since this term is already of ${\cal{O}}(T)$, as already seen in the clean case Eq.\ \pref{jdc_app}.  
By means of the same formalism used so far it is easy to verify that the first temperature correction reads:

\be
\begin{split}
&-\frac{1}{2}\overline{\braket{ \int d\mathbf{r} (\tilde{J} \Big(\nabla_x \theta(\mathbf{r})\Big)^2 }}_{H_0 + H^{(2)}_i}=\\&= - \frac{T}{2d}\Big[ 1 + \sum_i P_i\sum_{n=1}^{\infty} (-1)^n \Big(\frac{\delta J_i}{d \tilde{J}}	\Big)^n \Big] = - \frac{T}{2d},
\end{split}
\ee
where in the last step we used Eq.\eqref{request}, which accounts for the vanishing of the local self-energy. For the same reason also the  second temperature correction vanishes:
\be
\begin{split}
&-\frac{1}{2}\overline{\braket{ \delta J_i \Big(\nabla_x \theta(\mathbf{r}_i)\Big)^2 }}_{H_0 + H_i^{(2)}}=\\ &= \frac{T}{2}[ \sum_i P_i\sum_{n=1}^{\infty} (-1)^n \Big(\frac{\delta J_i}{d \tilde{J}}	\Big)^n \Big] = 0.
\end{split}
\ee
Hence at all orders in the perturbative expansion, the diamagnetic response function, up to the linear terms in temperature, only depends on the mean value of the random couplings and on the dimension of the system:

\be
J_d(T)= \bar{J}- \frac{T}{2d},
\label{Jd_T}
\ee
\noindent
showing a very remarkable universality with respect to the random couplings distribution itself. Finally, the temperature dependence of the paramagnetic term $J_p(T)= J_d(T) -\tilde J(T)$ can be obtained by combining the results Eq.\eqref{new_EMT} and  Eq.\eqref{Jd_T} for for $\tilde J(T)$ and for $J_d(T)$, respectively.

\section{Comparison with the Monte Carlo results}

In this  section, we will compare the analytical results, previously derived, with the numerical solutions obtained by means of Monte Carlo simulations on the classical XY model in the presence of random, spatially-uncorrelated, couplings $J^{\mu}_i$. 
The Monte Carlo simulations have been performed on systems with linear size $L=128$ and periodic boundary conditions. Each Monte Carlo step consists of five Metropolis spin flips of the whole lattice, needed to probe the correct canonical distribution of the system, followed by ten Over-relaxation sweeps of all the spins, which help the thermalization. For each temperature we perform 5000 Monte Carlo steps, and we compute a given quantity  averaging over the last 3000 steps,  discarding  thus the transient regime which occurs in the first 2000 steps. Furthermore, the thermalization at low temperatures is speeded up by a temperature annealing procedure. Finally, the average over disorder is done on 15 independent configurations for each disorder level considered. Where not shown, the error bars are smaller than the point size. 

We will consider two different disorder distributions for the couplings $J^{\mu}_i$, showing that in both cases the EMT equations  previously obtained are in very good agreement with the numerical results. We first consider the case of a Gaussian distribution:
\be
P(J^{\mu}_i) = \frac{1}{\sqrt{2 \pi \sigma}} \exp\Big[{-\frac{(J^{\mu}_{i}-J)^2}{2\sigma^2}}\Big],
\ee
where $J$ is set equal to one and the standard deviation $\sigma$ measures the disorder strength. We also consider the additional constraint of $J^{\mu}_i\geq0$ to prevent the presence of antiferromagnetic couplings. The zero-temperature value of the stiffness can be easily obtained by means of the explicit expressions for the diamagnetic and paramagnetic contributions derived in Sec. IIA. Indeed, by using Eq.\ \pref{Jd_T0} and Eq.\ \pref{jp0} we can easily estimate 
\bea
\label{Gauss_Jd0}
J_d(T=0)&=& \bar{J}\\
\label{Gauss_Jp0}
J_p(T=0)&=& \frac{\sigma_J^2}{d \bar{J}}\\
\tilde{J}(T=0)&=& \bar{J}\Big[ 1 - \frac{\sigma_J^2}{d \bar{J}^2}\Big]
\label{Gauss_Js0}
\eea
where for the paramagnetic term we just retained the leading term in $\delta J_i$, as given by Eq.\  \pref{jplead}.
 Eq. \pref{Gauss_Js0} could also be obtained\cite{stroud_prb00} by directly solving the EMA Eq.\ \pref{EMT} at leading order in $\delta J_i$. At finite temperature, we will follow indeed this procedure, starting from the self-consistency Eq.\ \eqref{new_EMT}. Since for the Gaussian distribution all the odd momenta are on average zero, it is convenient to express  $J_i= \bar{J}+ \Delta J_i$. We can then rewrite Eq.\ \eqref{new_EMT} as:
\be
\sum_i P_i \Big\{ \frac{\bar{J}+ \Delta J_i-\tilde{J} - \frac{T}{2}\frac{\bar{J}+ \Delta J_i}{(d-1)\tilde{J} +\bar{J}+ \Delta J_i}}{d\tilde{J} + \bar{J}+ \Delta J_i -\tilde{J} - \frac{T}{2}\frac{\bar{J}+ \Delta J_i}{(d-1)\tilde{J} +\bar{J}+ \Delta J_i}}\Big\}=0.
\ee
By retaining all the terms of orders ${\cal{O}}(T)$ and ${\cal{O}}((\Delta J_i)^2)$, after simple algebra we obtain:
\be
\tilde{J}\Big( 1 + \frac{T}{2d^2\bar{J}}\Big) \simeq \bar{J}( 1 + \frac{T}{2d^2\bar{J}}\Big) - \frac{\sigma_J^2}{d\tilde{J}} -\frac{T}{2d},
\ee
so that the effective stiffness reads:
\be
\tilde{J}\simeq \bar{J}\Big[ 1 - \frac{\sigma_J^2}{d \bar{J}^2}\Big] - \frac{T}{2d}.
\label{Gauss_JsT}
\ee
Finally, by using the general result Eq.\eqref{Jd_T} for the linear temperature dependence of the diamagnetic term, we can derive the separate expressions for both $J_d(T)$ and $J_p(T)$:
\bea
\label{Gauss_Jd}
J_d(T)&\simeq & \bar{J}- \frac{T}{2d} \\
\label{Gauss_Jp}
J_p(T)& \simeq & \frac{\sigma_J^2}{d \bar{J}} 
\eea

\begin{figure}[b!]
\centering
\includegraphics[width=0.82\linewidth]{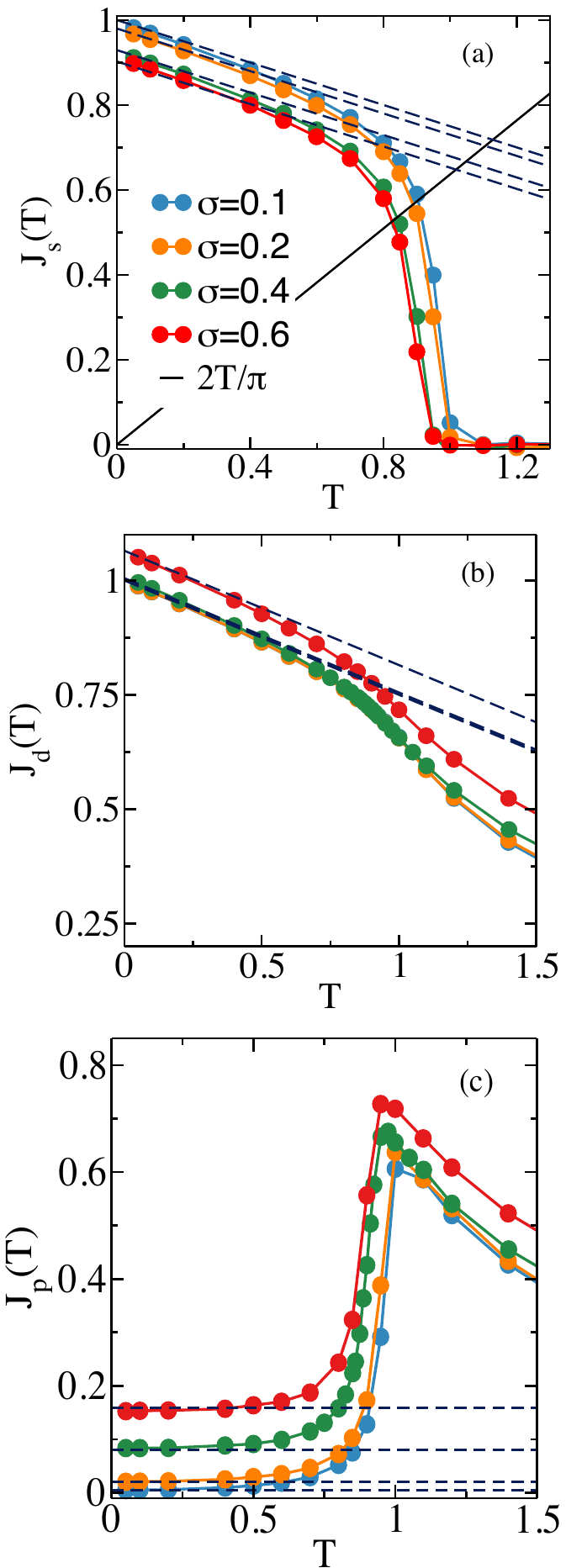}
\caption{(Color online) Monte Carlo results on a disordered XY model with gaussian distributed random couplings. In the three panels are plotted the three response functions: (a) $J_s$, (b) $J_d$ and (c) $J_p$ as function of the temperature. The dashed lines correspond to the analytical results \pref{Gauss_JsT}-\pref{Gauss_Jp} obtained from the EMA.=, and reproduces perfectly the numerical simulations.}
\label{fit_gauss}
\end{figure}
\noindent
\noindent
Eq.s\ \pref{Gauss_JsT}-\pref{Gauss_Jp} can be now compared with the numerical simulations. In Fig.\ref{fit_gauss}, we can see that for all the value of $\sigma$, the analytical  results fit very well the Monte Carlo data. Notice that for the case of $\sigma=0.6$ the truncation at $J^{\mu}_i\geq 0$ shifts the mean value of the couplings to $\bar{J}_{\sigma=0.6}\simeq 1.06 \, J >J$. Nevertheless, by using this value in Eq.\eqref{Gauss_JsT}-\eqref{Gauss_Jp}, we can perfectly reproduce the numerical results.

\begin{figure}[b!]
\centering
\includegraphics[width=0.85\linewidth]{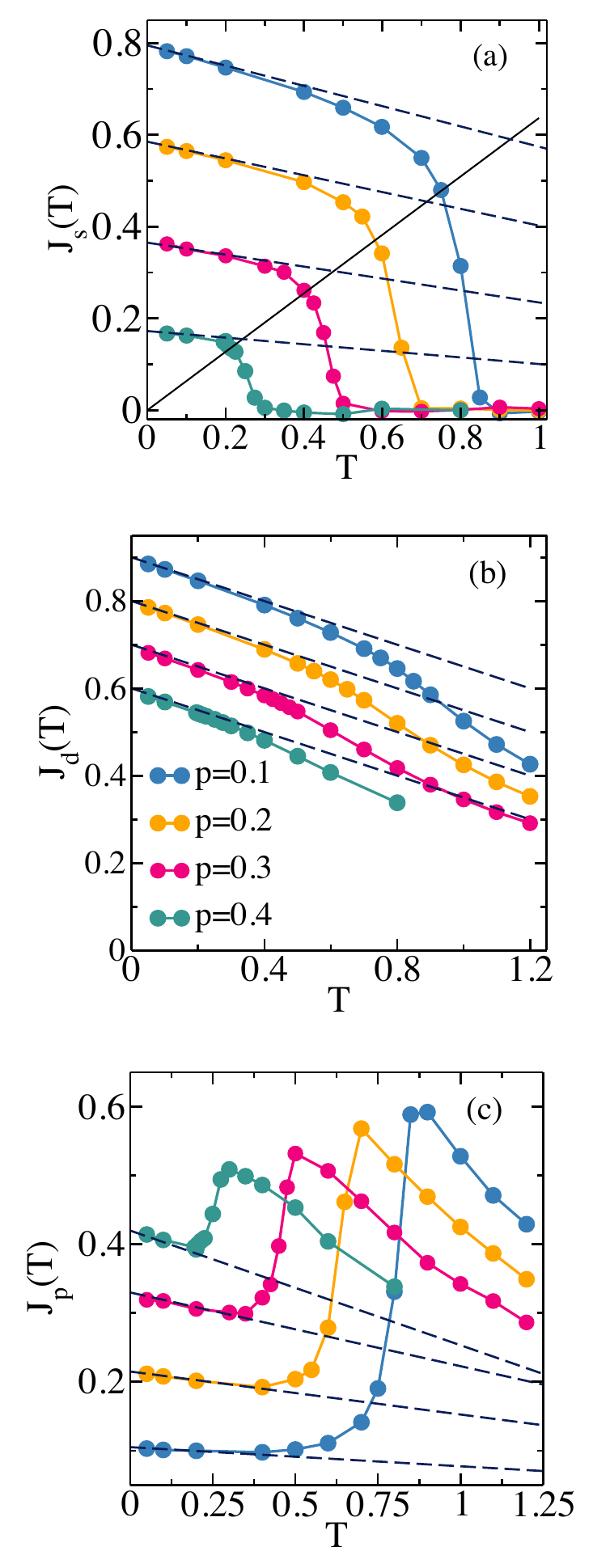}
\caption{(Color online) Monte Carlo results on a disordered XY model with random diluted couplings. In the three panels are plotted the three response functions: (a) $J_s$, (b) $J_d$ and (c) $J_p$ as function of the temperature. The dashed lines correspond to the analytical results \pref{fit_js}-\pref{fit_jp} obtained from the EMA. Even close to the percolation threshold at $p=0.5$, above which no superfluidity is possible, the analytical formula fit very well the numerical data.}
\label{fit_dil}

\end{figure}
The second kind of spatially uncorrelated disorder investigated is that of random diluted couplings, whose probability distribution reads:
\be
P(J^{\mu}_i)= p\, \delta(J^{\mu}_i) + (1-p)\,\delta(J^{\mu}_i -J)
\label{Prob_dil}
\ee
where $\delta(x)$  is the Dirac delta function and the dilution parameter $p$ will be the measure of the disorder strength. Also in this case we will set $J=1$ in the simulations. 
In this case, we cannot expect that the leading expansion in $\delta J_i$ holds, since for the diluted model the variance of the distribution  is of the same order of its average value. However, in this case the EMA equation can be explicitly solved. By considering directly the finite-temperature case Eq. \eqref{new_EMT} reads:
\be
\begin{split}
&\sum_i P_i \frac{-\delta J_i(T)}{d\tilde{J} +\delta J_i(T)}=\\ &= (1-p) \frac{ \tilde{J} -J + \frac{T}{2} \frac{1}{(d-1)\tilde{J} + J}}{(d-1)\tilde{J}  + J- \frac{T}{2} \frac{1}{(d-1)\tilde{J} + J}} + p =0,
\end{split}
\ee
\noindent
which, after some math, leads to: 

\be
\tilde{J}= J(1-2p)\Big( 1 -\frac{T}{2(J+\tilde{J})}\Big) \simeq J(1-2p)\Big( 1 - \frac{T}{4J(1-p)}\Big),
\ee
where in the last step we replaced $\tilde{J}$ in the denominator with its zero-temperature value. Using again Eq.\eqref{Jd_T} for the diamagnetic term, with $\bar{J}=J(1-p)$, we have the final expressions for all the three contributions:

\bea
J_s&=& J(1-2p) \Big[ 1 - \frac{T}{4J(1-p)}\Big]=\nn\\
\label{fit_js}
&=& J_s(T=0) \Big[ 1 - \frac{T}{4J(1-p)}\Big],\\
\label{fit_jd}
J_d&=& J(1-p) - \frac{T}{4}= J_d(T=0) -\frac{T}{4},\\
\label{fit_jp}
J_p&=& Jp - \frac{T}{4}\frac{p}{1-p}= J_p(T=0)\Big[1 -\frac{T}{4J(1-p)}\Big].
\eea

\noindent
In Fig.\ref{fit_dil} we show the Monte Carlo results for the three separate contributions. As one can see, the zero-temperature value of $J_p$ matches reasonably well the analytical estimate for $p=0.1$, while it slightly increases for larger disorder. Nonetheless, once accounted for this small deviation the temperature dependence still follows Eq.\ \pref{fit_jp},  as one can see by looking at the dashed lines in the three panels. In contrast, $J_d(T)$ follows closely the analytical estimate \pref{fit_jd} up to the largest disorder level. This suggests that, in analogy with what mentioned before for the thermal  effects due to vortices, also thermal effects due to longitudinal phase fluctuations affect more strongly the paramagnetic contribution than the diamagnetic one. This is presumably due to the long-range character of the correlations probed by the current-current response function, as compared to the local nature of the diamagnetic response. As a consequence, while for Gaussian distribution the leading temperature effects due to longitudinal phase fluctuations can be solely ascribed to the diamagnetic correction, for the diluted model the paramagnetic response is responsible for a substantial flattening of the thermal suppression of the superfluid stiffness, causing in turn an increase of the $T_c/J_s(0)$ ratio. 

\section{Conclusions}

In summary, we analyzed the effect of spatially-uncorrelated random disorder on the low-temperature behavior of the superfluid stiffness of the 2D classical XY model. We employed a perturbative expansion in the disorder potential that is analogous to the usual $T$-matrix scheme including only non-crossing diagrams. We found that this approach allows one to derive a self-consistency equation for the global stiffness that is fully equivalent to the usual EMA, usually discussed within the context of the RRN model\cite{kirkpatrick}. This result leads to two interesting consequences. First, it allows one to incorporate also the  finite-temperature corrections. This leads to the modified EMA equation \pref{new_EMT} for the stiffness, that properly describes the thermal suppression of the stiffness due to longitudinal phase modes in the presence of disorder. Second, it allows one to compute separately the diamagnetic and paramagnetic contributions to the stiffness. This is in turn a crucial information in order to establish the fraction of the total SC spectral weight which is transferred, thanks to the optical sum rule\cite{stroud_prb00}, to the finite-frequency absorption. These analytical findings offer an excellent description of the Monte Carlo results for both the Gaussian and diluted model of disorder. In the latter case the only discrepancy is a slightly larger paramagnetic suppression of the stiffness at $T=0$ for large disorder, that can be presumably ascribed to emerging space-correlation effects, neglected by the $T$-matrix scheme.  However, it is interesting to note that the resulting temperature suppression of the stiffness turns out to be {\em weaker} in the diluted model with respect to the homogeneous case. This result has to be contrasted with the recent Monte Carlo simulations\cite{BroadeningBKT} done with a granular space-correlated  model of disorder. In this case the {\em stronger}  thermal suppression of the stiffness with respect to homogeneous case had been attributed to a low-temperature proliferation of vortex-antivortex pairs in the bad SC regions. These opposite trends suggest that disorder not only affects the temperature scales where longitudinal (spin-wave like) and transverse (vortex-like) phase fluctuations become visible, but it can also profoundly change their interplay. Understanding how this interplay evolves as a function of disorder, and how it can be relevant for 2D SC systems, is an open question for future work.

\section{Acknowledgements}
This work has been supported  by Italian MAECI under the Italian-India
collaborative  project  SUPERTOP-PGR04879.  We acknowledge the cost action Nanocohybri CA16218.

\end{document}